\documentclass[10pt,twocolumn]{article}
\pdfoutput=1

\usepackage[]{graphicx}
\usepackage{amssymb,amsfonts,amsmath}
\usepackage{cite}
\usepackage[top=1in, bottom=1in, left=0.7in, right=0.7in]{geometry}
\usepackage{authblk}
\usepackage[font=small,labelfont=bf]{caption}
\usepackage{abstract}

\usepackage{titlesec}
\titleformat*{\subsection}{\normalsize\bfseries}

\begin{document}

\title{\textbf{Effective harmonic potentials: insights into the internal cooperativity and sequence-specificity of protein dynamics}}
\author{Yves Dehouck$^{ 1,2}$ and Alexander S. Mikhailov$^{ 1}$}
\affil{$^{1 }$Department of Physical Chemistry, Fritz-Haber-Institut der Max-Planck-Gesellschaft, Faradayweg 4-6, 11495 Berlin, Germany.}
\affil{$^{2 }$Department of BioModelling, BioInformatics and BioProcesses, Universit\'e Libre de Bruxelles (ULB), CP165/61, Av. Fr. Roosevelt 50, 1050 Brussels, Belgium.}
\date{}

\twocolumn[
\begin{@twocolumnfalse}
\maketitle
\begin{center}
{\tt ydehouck@ulb.ac.be} $\qquad \qquad \qquad$  {\tt mikhailov@fhi-berlin.mpg.de}
\end{center}
\vspace*{0.5cm}
\begin{abstract}
The proper biological functioning of proteins often relies on the occurrence of coordinated fluctuations around their native structure,
or of wider and sometimes highly elaborated motions.
Coarse-grained elastic-network descriptions are known to capture essential aspects of conformational dynamics in proteins,
but have so far remained mostly phenomenological, and unable to account for the chemical specificities of amino acids.
Here, we propose a method to derive residue- and distance-specific effective harmonic potentials
from the statistical analysis of an extensive dataset of NMR conformational ensembles.
These potentials constitute dynamical counterparts to the mean-force statistical potentials commonly used for static analyses of protein structures.
In the context of the elastic network model, they yield a strongly improved description of the cooperative aspects of residue motions,
and give the opportunity to systematically explore the influence of sequence details on protein dynamics.
\end{abstract}
\vspace*{1cm}
\end{@twocolumnfalse}
]


\section*{Introduction}

Deciphering the motions that underlie many aspects of protein function is a major current challenge in molecular biology,
with the potential to generate numerous applications in biomedical research and biotechnology.
Although molecular dynamics (MD)
hold a prominent position among computational approaches, considerable efforts have been devoted to
the development of coarse-grained models of protein dynamics \cite{Takada12}. Besides their ability to follow motions
on time scales that are usually not accessible to MD simulations, these models also give the possibility
to better understand the general principles that rule the dynamical properties of proteins. 13 \AA.

The elegant simplicity of the elastic network models (ENM) certainly contributed to their popularity,
and they have been successfully exploited in a wide range of applications \cite{Bahar10a,Bahar10b}. In these models, the residues
are usually represented as single particles and connected to their neighbors by Hookean springs \cite{Thirion96,Atligan01}.
The input structure is assumed to be the equilibrium state, i.e. the global energy minimum of the system.
Common variants include the homogeneous ENM, in which springs of equal stiffness connect pairs of residues separated by a distance smaller than a predefined cutoff,
and other versions in which the spring stiffness decays as the interresidue
distance increases \cite{Hinsen00,Moritsugu07,Yang09}. In all cases, the equations of motion can be either linearized around equilibrium,
to perform a normal mode analysis of the system \cite{Bahar05,Wynsberghe06,Dykeman10},
or integrated to obtain time-resolved relaxation trajectories \cite{Flechsig10,Duttmann12}. 

Despite their many achievements, purely structural ENM also come with severe limitations.
Notably, modeling the possible effects of mutations within this framework usually requires random local perturbations
of the spring constants \cite{Zheng05}, or a more drastic removal of links from the network \cite{Hamacher08}. A few attempts have
been made to include sequence-specificity in the ENM by setting the spring constants proportional to the depth
of the energy minima, as estimated by statistical contact potentials \cite{Hamacher06,Gerek09}. However, this approach cannot be
extended to distance-dependent potentials, for they are not consistent with the ground
hypothesis of the ENM, i.e. that all pairwise interaction potentials are at their minimum in the
native structure. Other studies have led to the conclusion that the ENM behave as entropic models dominated by
structural features, and that the level of coarse-graining is probably too high to incorporate sequence details \cite{Atligan01,Lezon10}.
Still, the chemical nature of residues at key positions can have significant effects on the main dynamical properties
of a protein. Hinge motions \cite{Gerstein98}, for instance, obviously require some architectural conditions to be fulfilled,
such as the presence of two domains capable of moving relatively independently. But the amplitude and preferred
direction of the motion are most likely determined by fine tuning of specific interactions
in the hinge region. In proteins subject to domain swapping, the hinge loops have indeed been shown to frequently
include residues that are not optimal for stability \cite{Dehouck03}. The importance of the amino acid sequence has also been
repeatedly emphasized by experimental studies of the impact of mutations on the conformational dynamics of proteins \cite{Siggers07,Trivedi12,Adhikary12}.

A major obstacle to the definition of accurate coarse-grained descriptions of protein dynamics lies in the highly cooperative nature of protein motions,
which makes it difficult to identify the properties of the individual building blocks independently of the overall architecture of each fold.
By condensing the information contained in a multitude of NMR ensembles, we build here a mean protein environment,
in which the behavior of residue pairs can be tracked independently of each protein's specific structure.
This methodology brings an efficient way of assessing coarse-grained models of protein dynamics
and of deriving effective energy functions adapted to these models.
In the context of the ENM, we identify a set of spring constants that depend on both the interresidue distances and the chemical nature of amino acids,
and that markedly improve the performances of the model.

\section*{Results}
\subsection*{Dynamical properties of proteins from the perspective of an average pair of residues}

The mean-square fluctuations of individual residues (MSRF) have been extensively relied
on to characterize protein flexibility and to evaluate coarse-grained models of protein dynamics \cite{Fuglebakk12},
in part because of their widespread availability as crystallographic B-factors. However, since the MSRF
carry little information about the cooperative nature of residue motions, we propose to examine the dynamical behavior of proteins
from the perspective of residue pairs rather than individual residues.
Information about the fluctuations of interresidue distances is contained in 
the data of NMR experiments for numerous proteins, and will be exploited here.
We define the apparent stiffness of a pair of residues $i$,$j$ in a protein $p$:

\begin{equation}
\gamma_{pij}=2k_{B}T/\sigma^2_{r_{pij}}
\end{equation}

\noindent
where $k_B$ is the Boltzmann constant, $T$ the temperature, and $\sigma^2_{r_{pij}}$ the variance of the distance $r$
between residues $i$ and $j$, in a structural ensemble representative of the equilibrium state.
$\gamma_{pij}$ is defined up to a multiplicative factor, which corresponds to the temperature.
We also introduce the uncorrelated apparent stiffness $\gamma\ensuremath{^\circ}_{pij}$,
to quantify the impact of the individual fluctuations of residues $i$ and $j$
on the fluctuations of the distance that separates them.
This is achieved by using $\sigma\ensuremath{^\circ}_{r_{pij}}$ instead of $\sigma_{r_{pij}}$ in eq. {\bf 1},
where $\sigma\ensuremath{^\circ}_{r_{pij}}$ is computed after exclusion of all correlations between the motions of residues $i$ and $j$
({\it Methods}).

\begin{figure}[t]
\centerline{\includegraphics[width=.48\textwidth]{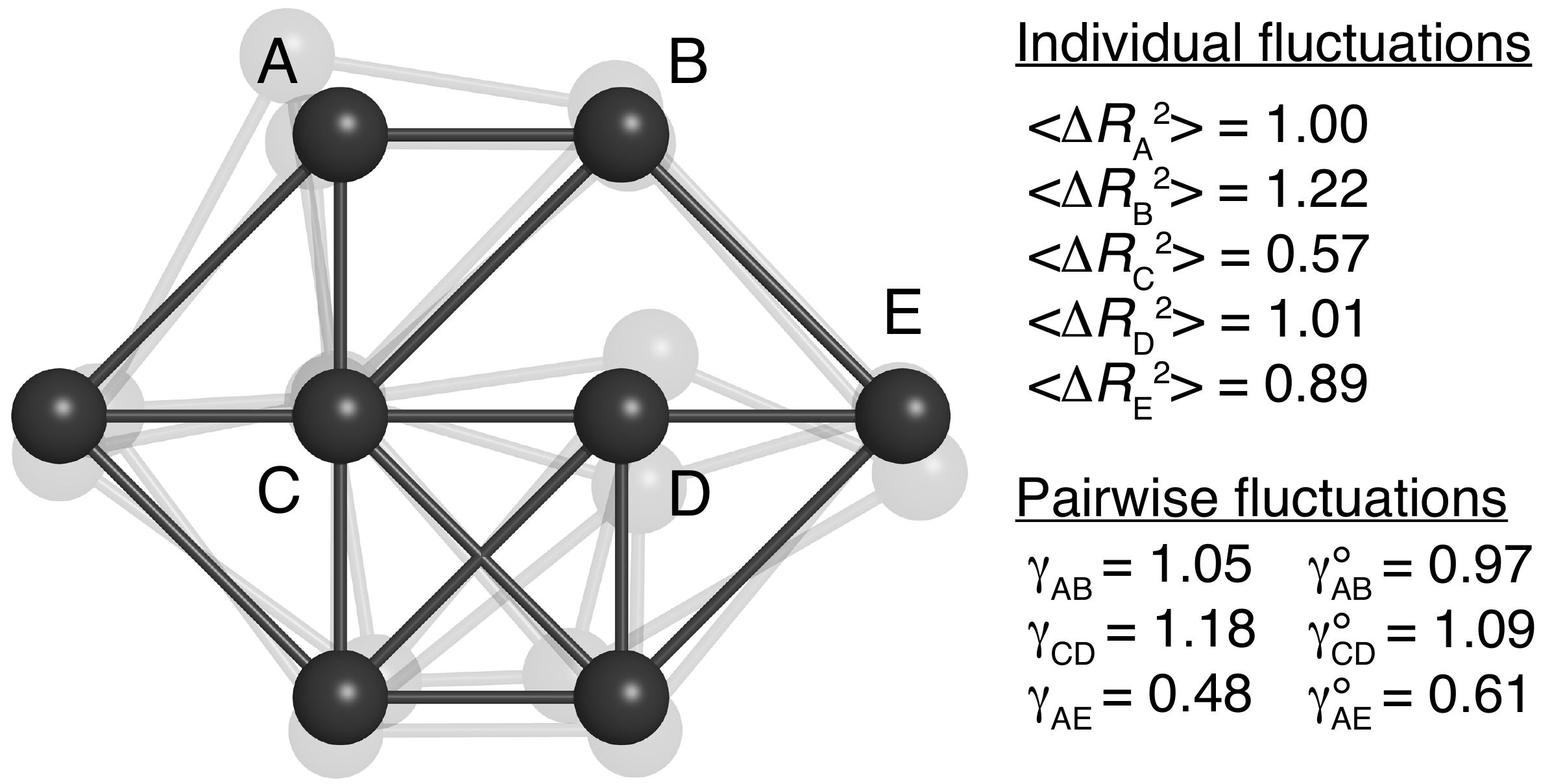}}
\caption{Schematic illustration of the apparent stiffness $\gamma$.
A simple model containing 8 beads connected by elastic springs was subjected to $10^7$ integration steps under Gaussian noise.
Selected values of $<(\Delta R_i)^2>$, $\gamma$ and $\gamma\ensuremath{^\circ}$ are given in arbitrary units.
Individually, the pairs A-B and C-D would be identical,
but they experience differently the influence of the other beads. As a result, the C-D pair is effectively more rigid
than A-B ($\gamma_{AB}<\gamma_{CD}$). In both cases, the motions are somewhat correlated, as the apparent stiffness $\gamma$ is larger than what
is expected from the knowledge of their individual motions ($\gamma\ensuremath{^\circ}$). Beads A and E do not interact directly but the effect
of the network on their relative motions is captured by the values of $\gamma_{AE}$ and $\gamma\ensuremath{^\circ}_{AE}$.}
\label{gamma}
\end{figure}

As illustrated in Figure 1, $\gamma$ can be quite different from one residue pair to another.
Indeed, besides the impact of direct interactions, $\gamma$ is also strongly dependent on the overall fold of the protein,
and on the position of the pair within the structure.
To remove the specific influence of each protein's architecture,
we define the apparent stiffness in a mean protein environment $\overline{\gamma}(s,d)$:

\begin{equation}
\overline{\gamma}(s,d)=\frac{2k_{B}T}{\sigma^2_r(s,d)}
\text{, with } \sigma^2_r(s,d)=\frac{\sum_{p=1}^P \sum_{ij}^{N_p(s,d)} M_p \sigma^2_{r_{pij}}}{\sum_{p=1}^P N_p(s,d) M_p}
\end{equation}

\noindent
where $s$ is one of 210 amino acid pairs, $d$ the discretized equilibrium distance between pairs of
residues ($d \leq r_{pij} < d+0.5\text{\AA}$), $M_p$ the number of structures in the equilibrium ensemble of protein $p$,
and $N_p(s,d)$ the number of $(s,d)$ residue pairs in protein $p$. Pairs of consecutive residues were dismissed,
so as to consider only non-bonded interactions.
The mean protein environment is thus obtained by
averaging over a large number of residue pairs in a dataset of $P=1500$ different proteins ({\it Methods}).

The influence of the distance separating two residues on the cooperativity of their motions can be investigated 
by considering amino acid types indistinctively in eq. {\bf 2}.
Interestingly, $\overline{\gamma}(d)$ follows approximately a power law, with an exponent of about -2.5 (Fig. 2).
Finer details include a first maximal value occurring for $C_{\alpha}$-$C_{\alpha}$ distances between 5 and 5.5 \AA,
i.e. the separation between hydrogen-bonded residues within regular secondary
structure elements, and a second around 9 \AA, which corresponds to indirect, second neighbor, interactions.
The high level of cooperativity in residue motions is well illustrated
by the comparison of $\overline{\gamma}(d)$ and its uncorrelated counterpart $\overline{\gamma}\ensuremath{^\circ}(d)$.
Indeed, these two functions would take identical values if the variability of the distance between
two residues could be explained solely by the extent of their individual fluctuations.
In a mean protein environment, however, $\overline{\gamma}(d)$ is about two orders of magnitude
larger than $\overline{\gamma}\ensuremath{^\circ}(d)$ at short-range, and the difference
remains quite important up to about 30-40 \AA.

\begin{figure}[t]
\centerline{\includegraphics[width=.45\textwidth]{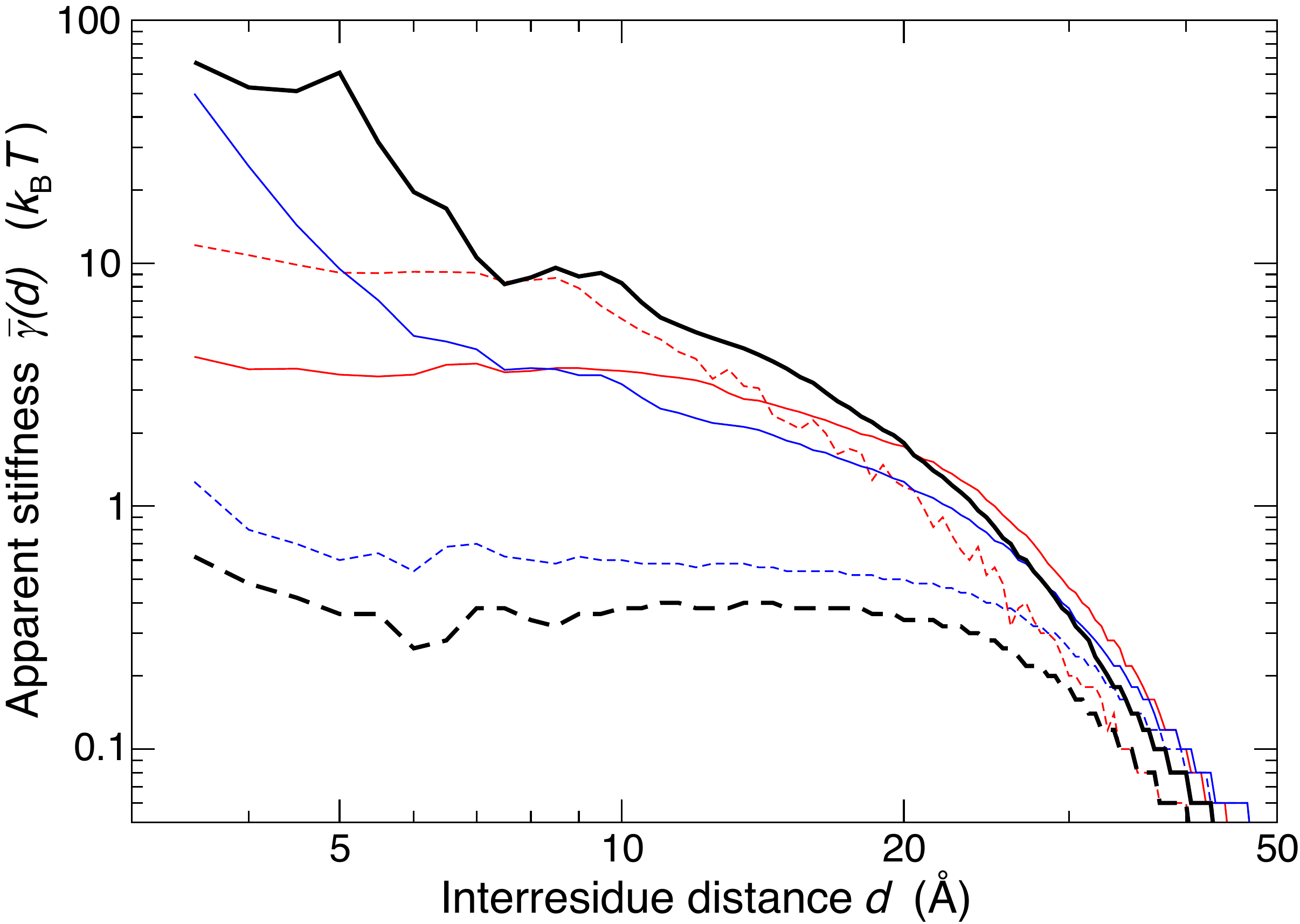}}
\caption{Comparison of the experimental and predicted values of the apparent stiffness $\overline{\gamma}(d)$.
Experimental values of $\overline{\gamma}(d)$ (continuous black) and $\overline{\gamma}\ensuremath{^\circ}(d)$ (dashed black),
extracted from the dataset of 1500 NMR ensembles.
Values of $\overline{\gamma}(d)$ predicted on the same dataset by the $\text{ENM}^0_{10}$ (dashed red); $\text{ENM}^0_{13}$ (continuous red);
$\text{ENM}^2_{50}$ (dashed blue); $\text{ENM}^6_{50}$ (continuous blue).
}
\label{gamma_d}
\end{figure}

The comparison of $\overline{\gamma}(d)$ values extracted from subsets containing exclusively small, large, all-$\alpha$,
or all-$\beta$ proteins indicates that the content of the dataset has a remarkably limited impact on $\overline{\gamma}(d)$ (Supplementary Fig. 1).
This distance dependence can thus be seen as a general property of protein structures, a signature of
protein cooperativity at the residue pair level. Of course, since $\overline{\gamma}(d)$ is representative of a mean
protein environment, deviations may occur for individual proteins,
according to their specific structural organizations (Supplementary Fig. 2).

The apparent stiffness $\overline{\gamma}(s)$ is computed for each type of amino acid pair $s$ using eq. {\bf 2},
by considering only residue pairs separated by less than 10 \AA.
As shown in Figure 3A, the chemical nature of the interacting residues is a major determinant of their dynamical behavior.
Unsurprisingly, Glycine and Proline appear as the most effective ingredients of flexibility.
Pairs involving hydrophobic and aromatic amino acids tend to be considerably more rigid,
with $\overline{\gamma}(s)$ values up to 6 times larger. These differences originate in part in
the individual propensities of different amino acids to be located in more or less flexible regions (e.g. hydrophobic core vs. exposed surface loops).
However, there is only a limited agreement between $\overline{\gamma}(s)$ and $\overline{\gamma}\ensuremath{^\circ}(s)$ (Fig. 3A-B):
the correlation coefficient is equal to 0.71, and $\overline{\gamma}(s)$ spans a much wider range of values.
Beyond individual amino acid preferences, the specifics of residue-residue interactions play thus
a significant role in determining the extent of cooperativity in residue motions.

\begin{figure*}[!t]
\centerline{\includegraphics[width=1.0\textwidth]{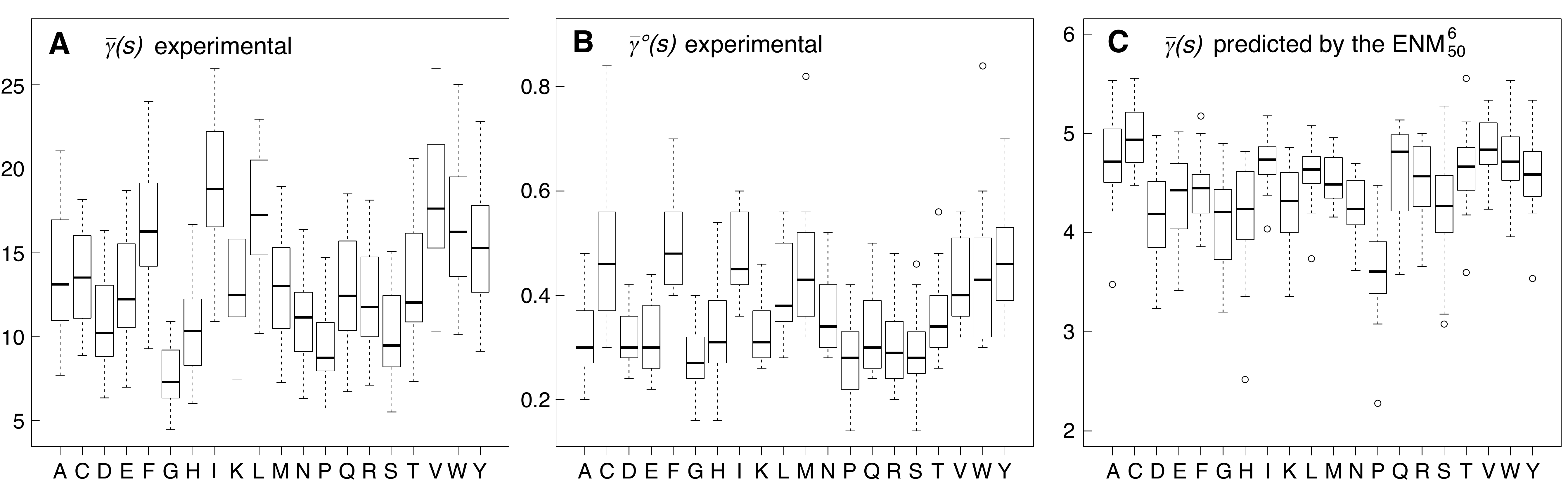}}
\caption{Comparison of the experimental and predicted values of the apparent stiffness $\overline{\gamma}(s)$, in the dataset of 1500 NMR ensembles.
For each amino acid, the median value of $\overline{\gamma}(s)$ over the 20 possible partners
is given in units of $k_BT$, along with the maximal, minimal, $1^{\text{st}}$ and $3^{\text{rd}}$ quartile values.
Outliers from these distributions are depicted as circles.
(A) Experimental values of $\overline{\gamma}(s)$.
(B) Experimental values of $\overline{\gamma}\ensuremath{^\circ}(s)$.
(C) Values of $\overline{\gamma}(s)$ predicted by the $\text{ENM}^6_{50}$.
}
\label{gamma_s}
\end{figure*}

\subsection*{Accuracy of elastic network models in reproducing the dynamical properties of proteins}

The computation of the apparent stiffness of residue pairs in a mean protein environment provides an interesting
tool to probe the dynamical properties of proteins. It also generates a
very straightforward approach to assess the ability of coarse-grained models to reproduce
accurately this general behavior.

We focus here on four common variants of the residue-based ENM \cite{Riccardi10,Leoiatts12},
which differ only by the functional form of the spring constants $\kappa$.
The dependence of $\kappa$ on the interresidue distance $r_{pij}$ is defined by
two parameters: the cutoff distance $l_d$, above which residues $i$ and $j$ are considered disconnected,
and the exponent $\alpha$ that determines how fast $\kappa$ decreases with increasing distances:

\begin{equation}
\kappa_{pij}(\text{ENM}^{\alpha}_{l_d})=a_p\mathcal{H}(l_d-r_{pij})r_{pij}^{-\alpha}
\end{equation}

\noindent
where $\mathcal{H}$ is the Heaviside function.
The value of the temperature-related factor $a_p$ is obtained, for each protein independently,
by fitting the predicted MSRF with the experimental ones.
This ensures that the amplitude of the individual fluctuations of the beads in the network is on average equal to that
observed in the corresponding NMR ensemble, and that the predicted $\overline{\gamma}(s,d)$ values can thus be directly
compared with those extracted from the NMR data. We consider the following models:
$\text{ENM}^{0}_{10}$, $\text{ENM}^{0}_{13}$, $\text{ENM}^{2}_{50}$, $\text{ENM}^{6}_{50}$.
These ENM variants were used to estimate the value of $\sigma^2_{r_{pij}}$ for each pair
of residues in the 1500 proteins of our NMR dataset (\textit{Methods}), and to subsequently compute
$\overline{\gamma}(d)$ and $\overline{\gamma}(s)$ from eq. {\bf 2}.

Strikingly, all ENM variants systematically predict $\overline{\gamma}(d)$ values to be lower than the experimental ones,
at least up to interresidue distances of 20-30 \AA (Fig. 2). These models overestimate thus the amplitude
of pairwise fluctuations, relatively to the amplitude of individual fluctuations.
For example, if two residues in a protein undergo highly correlated motions,
the amount of thermal energy necessary to induce a moderate variance on the distance between them
will generate high variances on their individual coordinates.
Consequently, if the motions of the beads of the ENM are less coordinated,
adjusting the scale of the spring constants to reproduce the amplitude of individual fluctuations
leads to an overestimated variance on the interresidue distances, and thus to lower $\overline{\gamma}(d)$ values.
This problem is particularly apparent when $\kappa$ is assumed to decrease proportionally to the square of the
interresidue distance, in the $\text{ENM}^{2}_{50}$.
Although this model was shown to perform well in predicting MSRF values \cite{Yang09},
our results suggest that it negates almost completely the coordinated aspect of
residue motions. Indeed, as shown in Figure 2, the $\overline{\gamma}(d)$ values predicted by this model are very
close to those obtained from the experimental data after removal of the correlations between the motions of
the different residues ($\overline{\gamma}\ensuremath{^\circ}(d)$). This observation is consistent with the extremely short atom-atom correlation
length characteristic of the $\text{ENM}^{2}_{50}$, recently estimated on the basis of an X-ray structure
of Staphylococcal nuclease \cite{Riccardi10}.

The ENM is often considered as an entropic model, not detailed enough to include sequence information
in a relevant way \cite{Atligan01,Lezon10}. It is therefore hardly surprising that common ENM variants produce a
poor description of the sequence specificities of protein dynamics. Individual amino acid preferences for more
or less densely connected regions are responsible for some variety in the predicted values of $\overline{\gamma}(s)$ (Fig. 3C).
However, this variety is far from matching the one observed in the experimental data,
as shown by a much narrower range of $\overline{\gamma}(s)$ values, and a limited correlation coefficient with the experimental
$\overline{\gamma}(s)$ values, e.g. 0.62 for the $\text{ENM}^{6}_{50}$ (Supplementary Fig. 3).

\subsection*{Derivation of effective harmonic potentials}

Mean-force statistical potentials are commonly used to perform energetic evaluations of static protein structures \cite{Sippl95,Miyazawa96,Dehouck06}.
These potentials do not describe explicitly the "true" physical interactions, but provide effective
energies of interaction in a mean protein environment, in the context of a more or less simplified
structural representation.
Similarly, within the ENM framework, $\kappa(s,d)$ defines for each pair of residues an harmonic interaction potential.
This potential is also effective in nature, accounting implicitly for
everything that is not included in the model (e.g. the surrounding water).
Hence, we seek to identify the value of $\kappa$ yielding the most accurate reproduction of
the dynamical behavior of each type of pair $(s,d)$ in a mean protein environment,
which is conveniently captured by the apparent stiffness $\overline{\gamma}(s,d)$.

\begin{figure*}[t]
\centerline{\includegraphics[width=0.95\textwidth]{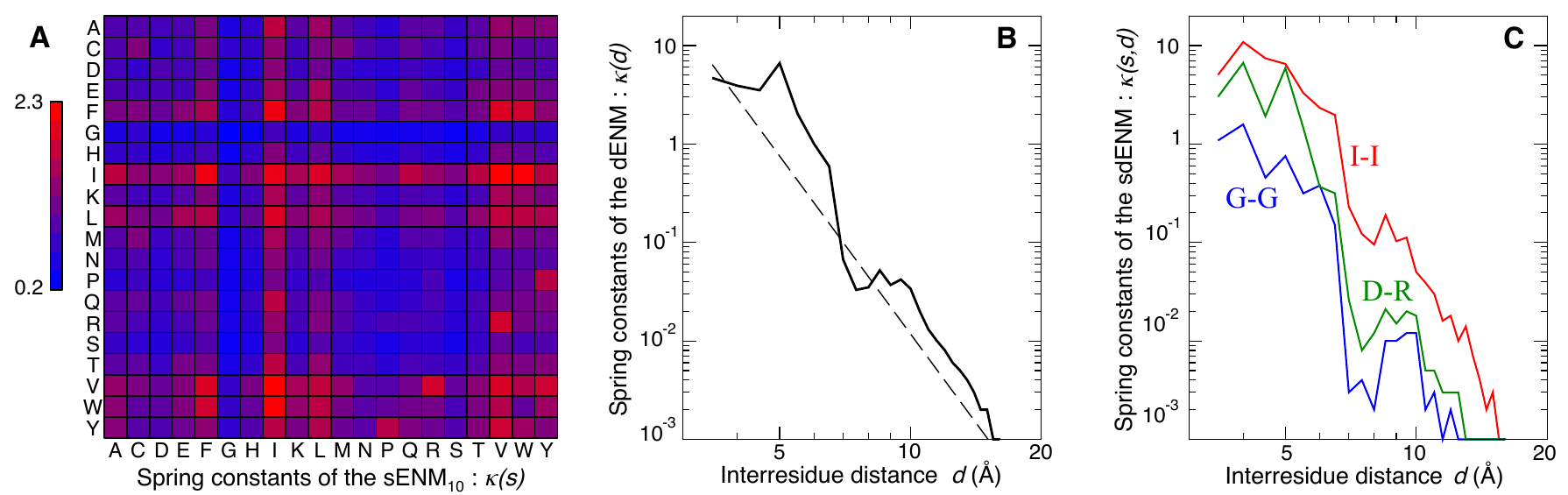}}
\caption{Effective harmonic potentials.
(A) Spring constants of the $\text{sENM}_{10}$, for the 210 amino acid pairs.
(B) Spring constants of the dENM. The dashed line corresponds to $\kappa \sim r^{-6}$.
(C) Spring constants of the sdENM for 3 amino-acid pairs.
All $\kappa$ values are given in Supplementary Tables 2-5.
}
\label{ehpot}
\end{figure*}

For that purpose, let us define $\overline{E}^{\text{bond}}(s,d)$ as the energy of the elastic spring connecting two residues of type $(s,d)$,
in a mean protein environment:

\begin{equation}
\overline{E}^{\text{bond}}(s,d)=\frac{1}{2}\kappa(s,d)\sigma^2_r(s,d)=k_BT\frac{\kappa(s,d)}{\overline{\gamma}(s,d)}
\end{equation}

\noindent
where $\overline{\gamma}(s,d)$ is the apparent stiffness extracted from the experimental data.
$\overline{E}^{\text{bond}}(s,d)$ is unknown and is expected to be different for different pair types $(s,d)$.
The knowledge of $\overline{\gamma}(s,d)$ is thus not sufficient to estimate directly $\kappa(s,d)$.
However, from any approximate set of spring constants $\kappa'(s,d)$, we may build the ENM for all proteins in our dataset,
to reproduce the mean protein environment, and compute for each pair type an estimated value of the apparent stiffness, $\overline{\gamma}'(s,d)$,
and bond energy, $\overline{E}'^{\text{ bond}}(s,d)$.

Since the behavior of a given residue pair is highly dependent on its environment, we can make the assumption that
$\overline{E}'^{\text{ bond}}(s,d)$ is a relatively good approximation of $\overline{E}^{\text{bond}}(s,d)$,
even if $\kappa'(s,d) \neq \kappa(s,d)$:

\begin{equation}
\overline{E}'^{\text{ bond}}(s,d)=k_BT\frac{\kappa'(s,d)}{\overline{\gamma}'(s,d)} \simeq \overline{E}^{\text{bond}}(s,d)
\end{equation}

\noindent
Indeed, if the spring stiffness of a residue pair is underestimated $(\kappa' < \kappa)$,
it will also appear as less rigid in the ENM than in the experimental data $(\overline{\gamma}' < \overline{\gamma})$.
A more detailed discussion is given in \textit{Supplementary Note} 1.
From eqs. {\bf 4} and {\bf 5}, we devise thus an iterative procedure in which $\kappa(s,d)$
is updated at each step $k$ by confronting the predicted values of the apparent stiffness, $\overline{\gamma}_k(s,d)$, with the experimental ones, $\overline{\gamma}(s,d)$.
It is expected to converge when $\overline{\gamma}_k(s,d) \rightarrow \overline{\gamma}(s,d)$,
that is, when the predictions of the model agree with the experimental data:

\begin{equation}
\kappa_{k+1}(s,d)=\kappa_{k}(s,d)\frac{\overline{\gamma}(s,d)}{\overline{\gamma}_k(s,d)}
\end{equation}

We used this approach to derive, from the NMR data, four novel ENM variants: the distance-dependent
dENM ; the sequence-dependent $\text{sENM}_{10}$ and $\text{sENM}_{13}$, with a distance cutoff of 10 and 13 \AA,
respectively, and the sequence- and distance-dependent sdENM (\textit{Methods}).
Interestingly, the $\kappa$ values for the 210 amino acid pairs in the $\text{sENM}_{10}$ are relatively well correlated
with the corresponding contact potentials \cite{Miyazawa96}, even though they result from totally different approaches (Supplementary Fig. 4).
Some common general trends can be identified,
e.g. hydrophobic contacts tend to be associated with both favorable interaction energies and large $\kappa$ values (Fig. 4A).
However, the overall correspondence remains limited, indicating that the determinants of protein rigidity and
stability are related, but distinct. The distance dependence of $\kappa$ in the dENM is remarkably similar to the $r^{-6}$
power law that was previously obtained by fitting against a 1.5ns MD trajectory of a C-phycocyanin dimer \cite{Hinsen00} (Fig. 4B),
although our new model presents more detailed features. Notably, $\kappa$ remains approximately constant up to
interresidue distances of 5-6 \AA, and then drops by about two orders of magnitude to reach a second plateau
between 7 and 12 \AA. The $\kappa$ values of the sdENM are shown in Figure 4C, for a few amino acid pairs.
This model not only combines the strengths of the sENM and the dENM, but also reveals the sequence specificity
of the $\kappa$ distance dependence. The D-R pair, for example, is almost as rigid as I-I at short distances consistent
with the formation of a salt bridge, but almost as flexible as G-G at larger distances.

\subsection*{Performances of the new ENM}

The sdENM yields a much more accurate reproduction of the dynamical behavior of residue pairs in a
mean protein environment than the common ENM variants, as demonstrated by the good agreement between
experimental and predicted values of $\overline{\gamma}(s)$ (Fig. 5A, Supplementary Fig. 5), and $\overline{\gamma}(d)$ (Fig. 5B).

\begin{figure*}[t]
\centerline{\includegraphics[width=1.0\textwidth]{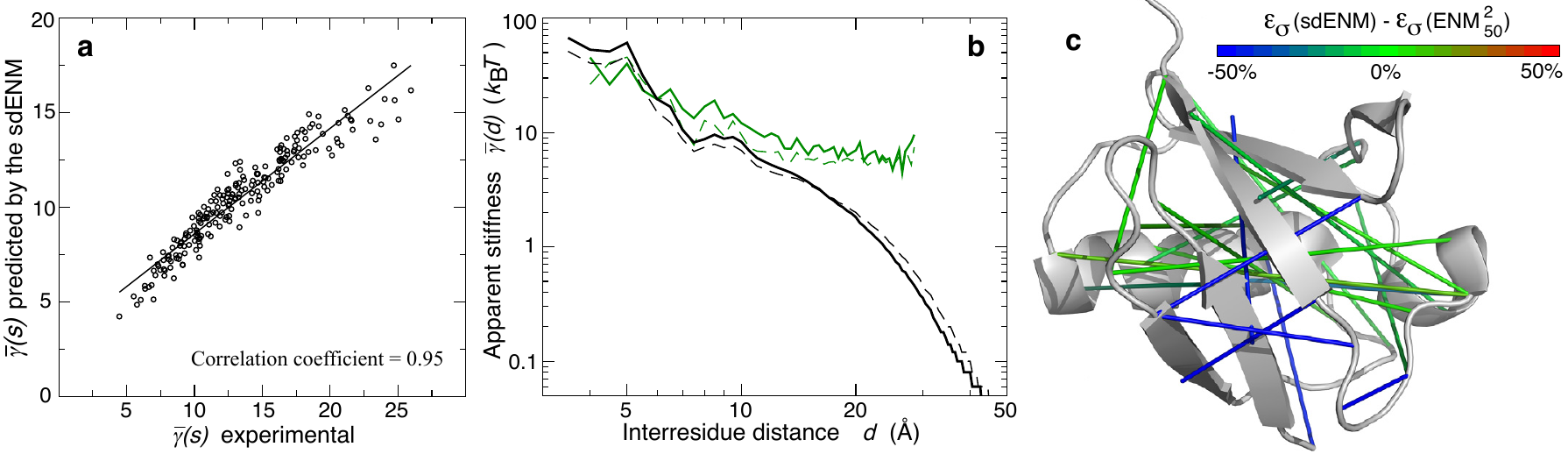}}
\caption{Performances of the sdENM.
(A) Experimental and predicted values of $\overline{\gamma}(s)$, in the dataset of 1500 NMR ensembles. See also Supplementary Figures 3 and 5.
(B) Experimental (continuous) and predicted (dashed) values of $\overline{\gamma}(d)$, in the dataset of 1500 NMR ensembles (black), 
and in a single protein (PDB 1xqq) (green). See also Supplementary Figure 2.
(C) Comparison of the ability of the sdENM and the $\text{ENM}^2_{50}$ to correctly reproduce the fluctuations of a single protein,
on the basis of a high quality structural ensemble of human ubiquitin (PDB 1xqq) \cite{Lindorff05}.
20 randomly selected residue pairs are connected by solid lines.
Shades of blue indicate a better performance of the sdENM, while shades of red indicate a better performance of the $\text{ENM}^2_{50}$.
See also Supplementary Figure 6.
}
\label{ehpot}
\end{figure*}

Beyond its performances in a mean protein environment,
our new model also brings highly notable improvements with respect to previously described ENM variants
when it is applied to the specific architecture of a given protein.
This is illustrated by two examples, on Figure 5C and Supplementary Figure 6.
A more thorough assessment of the ability of the different ENM variants to capture the motions of individual proteins
was performed on an independent dataset of 349 proteins. The correlation coefficient between predicted and
observed MSRF ($r_B$) has been widely used in the past but ignores the cooperativity inherent to protein dynamics,
and presents other shortcomings. Therefore, we introduce a new measure ($\epsilon_{\sigma}$) that quantifies the relative
error on the estimation of the variability of the distance between residue pairs, and is thus focused
on the cooperative aspects of residue motions (\textit{Methods}).

\begin{table}[b!]
\centering
\caption{Performances of different ENM variants. $^{(a)}$ Average correlation coefficient between experimental and measured MSRF. $^{(b)}$ Average relative error on the fluctuations of interresidue distances.}
\begin{tabular*}{0.48\textwidth}{@{\extracolsep{\fill}}lccccc}

	& $r_B^{ (a)}$ 
& $\epsilon_{\sigma}^{ (b)}$ 
& $\epsilon^{\text{SR}}_{\sigma}$ & $\epsilon^{\text{MR}}_{\sigma}$ & $\epsilon^{\text{LR}}_{\sigma}$ \\
\hline
$\text{ENM}^0_{10}$	& 0.63 & 0.59 & 0.53 & 0.59 & 0.68 \\
$\text{ENM}^0_{13}$	& 0.65 & 0.68 & 0.69 & 0.68 & 0.68 \\
$\text{ENM}^2_{50}$	& 0.66 & 0.97 & 1.07 & 0.96 & 0.74 \\
$\text{ENM}^6_{50}$	& 0.69 & 0.64 & 0.59 & 0.66 & 0.66 \\
$\text{sENM}_{10}$	& 0.63 & 0.55 & 0.49 & 0.55 & 0.67 \\
$\text{sENM}_{13}$	& 0.66 & 0.63 & 0.63 & 0.63 & 0.67 \\
$\text{dENM}$	    & 0.69 & 0.54 & 0.48 & 0.56 & 0.60 \\
$\text{sdENM}$	    & 0.70 & 0.48 & 0.41 & 0.49 & 0.57 \\
\hline
\label{table1}
\end{tabular*}
\end{table}

Among the 4 previously described ENM variants, $\text{ENM}^{6}_{50}$ is better at predicting the individual residue fluctuations (Table 1).
Interestingly, the $\text{ENM}^{0}_{10}$, with its simple cutoff distance, appears superior when
it comes to the reproduction of cooperative motions ($\epsilon_{\sigma}=0.59$).
The new ENM variants based on our effective harmonic potentials present
enhanced performances in comparison with the common models.
In particular, the dENM reaches the same level of quality as the
$\text{ENM}^{6}_{50}$ for individual fluctuations ($r_B=0.69$),
but surpasses even the $\text{ENM}^{0}_{10}$ for the description of cooperativity ($\epsilon_{\sigma}=0.54$).
On the other hand, the impact of introducing sequence specificity can be examined by comparing
$\text{sENM}_{10/13}$ with $\text{ENM}^{0}_{10/13}$, and sdENM with dENM.
It consists in a slight improvement of the correlation coefficient $r_B$, and a pronounced decrease of the error $\epsilon_{\sigma}$,
especially at short- (0-15 \AA) and mid- (15-30 \AA) range.

\section*{Discussion}

For the last decades, statistical potentials extracted from datasets of known protein structures \cite{Sippl95,Miyazawa96,Dehouck06} have played a
critical role in static analyses of protein structures, with major applications including structure prediction,
protein-protein docking, or rational mutant design. Our study demonstrates that a similar approach can be taken
to derive effective energy functions that are specifically adapted to the coarse-grained modeling of protein dynamics.

More precisely, in the context of the ENM, we exploited a dataset of 1500 NMR ensembles to determine the values
of the spring constants that describe best the behavior of pairs of residues, as a function of both their chemical
nature and the distance separating them. The success of our approach is attested by a drastic enhancement of the
ability to accurately describe the cooperative nature of residue motions, with respect to previously described ENM variants.
Moreover, a definite advantage of our method is that the effective parameters characterizing the strength of the virtual bonds
are directly extracted from the experimental data without any a priori conception of their functional form.
The fact that the distance dependence of the spring constants that we retrieve is quite similar to the $r^{-6}$ power law,
which was considered so far as underlying one of the best performing ENM variants \cite{Hinsen00,Riccardi10},
also constitutes a major support to our approach.

In our derivation scheme, the virtual bonds are parametrized so as to reproduce the behavior of amino acid pairs
in a mean protein environment.
The analysis of the ability of different models of protein dynamics to describe the motions of residues within this environment
sheds an interesting new light on the properties of these models. In particular,
our results indicate that previous ENM variants underestimate, sometimes dramatically, the rigidity of amino acid
pairs at short- and mid-range. Our new model does however provide a much more accurate reproduction of the balance
between short-range and long-range coordinated motions. This is arguably a crucial aspect when considering, for example,
the consequences of localized alterations induced by ligand binding on signal transduction or global conformational changes,
such as in ATP-powered molecular motors.

Importantly, our results also demonstrate that the ENM does not have to be exclusively structural,
and that sequence details may be allowed to play a major role in coarse-grained descriptions of protein dynamics.
Thereby, this study paves the way towards comparative analyses of motions in proteins that share a similar structure
but present differences in sequence. Such investigations will prove particularly interesting in the context of the
rational design of (modified) proteins with controlled dynamical properties. Although we focused here
on residue-based elastic network models, our approach is not limited to this particular family, and can be readily
implemented to evaluate and optimize other coarse-grained models of protein dynamics. Notably, the impact of chemical specificity
on the dynamical behavior of residues should be even more accurately rendered by effective potentials based
on a more detailed structural description.

\section*{Methods}
\subsection*{NMR Dataset}
We retrieved, from the Protein Data Bank \cite{Berman00}, ensembles of at least 20 models from solution NMR experiments,
corresponding to monomeric proteins of at least 50 residues that present at most 30\% sequence identity with one another.
Entries under the SCOP classifications "Peptides" or "Membrane and cell surface proteins" were not considered.
The presence of ligands, DNA or RNA molecules, chain breaks, non-natural amino acids, and differences in the number
of residues per model were also grounds for rejection. These criteria led to the selection of 1849 distinct structural ensembles.
A subset of 1500 ensembles was randomly selected for the main analysis, and the remaining 349 were used to assess the performances
of the different ENM variants. Unfolded C- or N-terminal tails were automatically identified
(MSRF values larger than twice the average for all residues in the protein) and removed from consideration.
In each ensemble, the structure with the lowest root mean square deviation from the mean structure, after superposition,
is chosen as representative and used to build the ENM.

\subsection*{Elastic network model}
The network is built by considering each residue as a single bead, placed at the position of the corresponding $C_{\alpha}$ atom
in the input structure, and connecting neighboring beads with Hookean springs \cite{Thirion96,Atligan01}. The ENM variants considered here differ only
by the form of the spring constant $\kappa$ as a function of interresidue distance and of amino acid types.
In all variants, bonded interactions are described by a larger value of $\kappa$, defined as ten times the value of $\kappa$
for non-bonded interactions at a separation of 3.5 \AA, averaged over all amino acid types.
The potential energy of the network is given by: $U=\sum_{i<j}(\kappa_{ij}/2)(r_{ij}-r\ensuremath{^\circ}_{ij})^2$,
where $r_{ij}$ and $r\ensuremath{^\circ}_{ij}$ are the instantaneous and equilibrium distances between residues $i$ and $j$,
respectively. By definition, the input structure corresponds to the global energy minimum, with $U=0$.
For a protein of $n$ residues, the Hessian $\mathbf{H}$ of the system is the $3n \times 3n$ matrix of
the second derivatives of $U$ with respect to the spatial coordinates of the residues.
The eigenvalue decomposition of $\mathbf{H}$ yields the covariance matrix $\mathbf{C}$ of the spatial coordinates,
which constitutes the output of the model:
\begin{equation}
\mathbf{C}=\sum_{k=1}^{3n-6}\frac{1}{\lambda_k}\mathbf{u}_k\mathbf{u}_k^\top
\end{equation}
\noindent
where the sum is performed over the $3n-6$ non-zero eigenvalues $\lambda_k$ of $\mathbf{H}$, and $\mathbf{u}_k$
are the corresponding eigenvectors. $\mathbf{C}$ is a $3n \times 3n$ symmetrical matrix, constituted of $n \times n$ submatrices $\mathbf{C}_{ij}$:
\begin{equation}
\mathbf{C}_{ij}=
	\begin{pmatrix}
	<\Delta x_i \Delta x_j> & <\Delta x_i \Delta y_j> & <\Delta x_i \Delta z_j> \\
	<\Delta y_i \Delta x_j> & <\Delta y_i \Delta y_j> & <\Delta y_i \Delta z_j> \\
	<\Delta z_i \Delta x_j> & <\Delta z_i \Delta y_j> & <\Delta z_i \Delta z_j>
	\end{pmatrix}
\end{equation}
\noindent
where $\Delta x_i$, $\Delta y_i$, and $\Delta z_i$ correspond to the displacements of residue $i$ from its equilibrium position,
along the three Cartesian coordinates. The predicted MSRF of residue $i$ is given by the trace of submatrix $\mathbf{C}_{ii}$:
$<(\Delta R_i)^2> = <(\Delta x_i)^2> + <(\Delta y_i)^2> + <(\Delta z_i)^2>$.

\subsection*{Variance of the interresidue distance}
For each pair of residues in a given protein $p$, the experimental value of this variance is readily computed from the NMR data:
\begin{equation}
\sigma^2_{r_{pij}}=\frac{1}{M—p} \sum_{m=1}^{M_p} (r_{pijm}-\overline{r}_{pij})^2
\end{equation}
\noindent
where $M_p$ is the number of structures in the NMR ensemble, $r_{pijm}$ the distance between the $C_{\alpha}$ atoms
of residues $i$ and $j$ in structure $m$ of protein $p$, and $\overline{r}_{pij}$ the average distance over all $M_p$ structures.
In the context of the ENM, $\sigma^2_{r_{pij}}$ values are estimated from the covariance matrix of the spatial coordinates,
by standard statistical propagation of uncertainty:
\begin{equation}
\sigma^2_{r_{pij}} \approx \mathbf{J}
	\begin{bmatrix}
	\mathbf{C}_{ii} & \mathbf{C}_{ij} \\
	\mathbf{C}_{ji} & \mathbf{C}_{jj}
	\end{bmatrix}
\mathbf{J}^\top
\end{equation}
\noindent
where $\mathbf{J}$ is the Jacobian of the distance $r_{pij}$ as a function of the six coordinates $(x_i,y_i,z_i,x_j,y_j,z_j)$.
This estimation of $\sigma^2_{r_{pij}}$ relies on the validity of the first order Taylor expansion of the distance as function
of the coordinates in the vicinity of the average distance. We ensured that no systematic bias arose from this approximation (Supplementary Fig. 7).
To quantify the impact of the individual motions of residues on their relative positions, we use eq. {\bf 10}
to compute $(\sigma\ensuremath{^\circ}_{r_{pij}})^2$ in an artificial construct where residue motions are not correlated.
This is achieved by extracting the covariance matrix from the NMR data, and setting to zero all submatrices $\mathbf{C}_{ij}$ where $i \neq j$.

\subsection*{Iterative procedure}

The values of the spring constants of the new ENM variants were derived from the dataset of 1500 NMR ensembles using eq {\bf 6}.
For the dENM, $\text{sENM}_{10}$ and $\text{sENM}_{13}$,
the initial values of the spring constants were set equal to the experimental values of the apparent stiffness:
$\kappa_0(d)=\overline{\gamma}(d)$ or $\kappa_0(s)=\overline{\gamma}(s)$.
Note that the $\overline{\gamma}(s)$ values were computed by considering only
residue pairs separated by a distance lower than the cutoff of 10 or 13 \AA.
For the sdENM, the $\kappa_0(s,d)$ values
were set equal to the final values of the spring constants in the dENM, $\kappa(d)$, for all amino acid types.
A correction for sparse data was devised to ensure that $\kappa(s,d)$ tends to $\kappa(d)$
when the number of residue pairs of type $(s,d)$ is too small to obtain relevant estimations of 
$\sigma^2_r(s,d)$.
Instead of eq. {\bf 2}, we used the following definition to compute both the experimental and predicted apparent stiffness:

\begin{equation}
\overline{\gamma}(s,d)= \frac{2k_{B}T}{\sigma^2_r(s,d)\left( \frac{N_{sd}}{N_{sd}+\mathcal{S}} \right)
+ \sigma^2_r(d)\left( \frac{\mathcal{S}}{N_{sd}+\mathcal{S}} \right)}
\end{equation}

\noindent
where $N_{sd}=\sum_{p=1}^P N_p(s,d) M_p$, $N_p(s,d)$ is the number of pairs of type $(s,d)$ in protein $p$,
$M_p$ is the number of structures in the NMR ensemble of protein $p$, and $\mathcal{S}$ is an adjustable
parameter set to 500.

The $\kappa$ values were rescaled after each iteration step, so that the average value of $\kappa$
over all amino acid types is equal to 1 for pairs separated by a distance of 6 \AA.
Residue pairs of a given type $(s,d)$ for which $\kappa(s,d)<0.001$ (after rescaling),
were considered to establish no direct interaction: $\kappa(s,d)$ was set to 0,
and they were no longer considered in the iterative procedure.
The performances of the new ENM variants after the first nine iteration steps are reported in Supplementary Table 1.
The procedure converged rapidly for the dENM and the sdENM, and the final models were selected after 5 and 3 iteration steps, respectively.
The sENM variants did not improve significantly with respect to the initial models ($k=0$), indicating that
$\kappa(s) = \overline{\gamma}(s)$ is a good approximation, contrary to $\kappa(d) = \overline{\gamma}(d)$.
The procedure was thus stopped after one iteration step, for both the $\text{sENM}_{10}$ and the $\text{sENM}_{13}$.

\subsection*{Performance measures}
The ability of coarse-grained models to accurately describe protein dynamics is commonly
evaluated by computing the Pearson correlation coefficient between predicted and experimental MSRF,
$<(\Delta R_i)^2>$, over all $i=1,...,n$ residues of a given protein:

\begin{equation}
r_B = \frac{\sum_{i=1}^n (B_i^{exp}-\overline{B})(B_i^{pre}-\overline{B})}
		{\sqrt{\sum_{i=1}^n (B_i^{exp}-\overline{B})^2}  \sqrt{\sum_{i=1}^n (B_i^{pred}-\overline{B})^2}}
\end{equation}

\noindent
where, for simplicity, $B_i$ was used instead of $<(\Delta R_i)^2>$.
There is indeed a direct relationship between the MSRF and the cristallographic B-factors:
$B_i=(8\pi^2/3) <(\Delta R_i)^2>$.
$B_i^{exp}$ and $B_i^{pre}$ correspond thus here to the MSRF of residue $i$ extracted
from the NMR data and predicted by the ENM, respectively.
The scale of the predicted MSRF values depends on the scale
of the spring constants, which are only defined up to a constant factor.
This factor was determined, for each protein independently,
by fitting the scales of the predicted and experimental MSRF, i.e. to ensure that:

\begin{equation}
\overline{B}=\frac{1}{n} \sum_{i=1}^n B_i^{exp} = \frac{1}{n} \sum_{i=1}^n B_i^{pre}
\end{equation}

\noindent
Although it has been widely used in previous studies, $r_B$ is probably not the most adequate measure
to evaluate the performances of coarse-grained models of protein dynamics.
As pointed out previously \cite{Riccardi10,Fuglebakk12}, it does indeed present several shortcomings:
e.g. it is strongly affected by the presence of highly flexible regions,
and does not account for possible flaws leading to an intercept of the regression line different from zero.
Most importantly, the MSRF describe individual fluctuations
but provide no information about the cooperative aspects of residue motions.
The quality of the MSRF predictions gives thus no guarantee
about the ability of the model to describe the cooperativity of protein dynamics. The $\text{ENM}_{50}^2$
provides an interesting example, for it performs quite well in predicting the MSRF but basically negates
all cooperativity (Fig. \ref{gamma_d}, Table \ref{table1}).

Therefore, we introduce a new measure
that exploits the information contained in the correlation matrix $\mathbf{C}$,
to quantify the error on the estimation of the fluctuations of the interresidue distances:

\begin{equation}
\epsilon_{\sigma}=\sqrt{\frac{1}{N_p} \sum_{ij}^{N_p} \left(\frac{\sigma^{\text{(exp)}}_{r_{pij}} - \sigma^{\text{(pre)}}_{r_{pij}}} {\sigma\ensuremath{^\circ}_{r_{pij}}}\right)^2}
\end{equation}

\noindent
where $N_p$ is the number of non-bonded residue pairs in protein $p$,
$\sigma^{\text{(exp)}}_{r_{pij}}$ and $\sigma^{\text{(pre)}}_{r_{pij}}$
are the experimental (eq. {\bf 9}) and predicted (eq. {\bf 10}) values of $\sigma_{r_{pij}}$, respectively.
$\sigma^{\text{(pre)}}_{r_{pij}}$ is obtained after fitting the experimental MSRF with the predicted ones (eq. {\bf 13}). 
The error is normalized by  $\sigma\ensuremath{^\circ}_{r_{pij}}$, which is
the expected value of $\sigma_{r_{pij}}$ given the individual, anisotropic, fluctuations of both residues extracted from
the NMR data, but neglecting all correlations between their respective motions.
This normalization ensures that the contributions of the different pairs of residues are equivalent,
and that the measure is not dominated by highly flexible regions.

Both $r_B$ and $\epsilon_{\sigma}$ are computed independently for each of the 349 proteins of our test set,
and the average values are reported. We also report the short- ($\epsilon^{\text{SR}}_{\sigma}$), mid- ($\epsilon^{\text{MR}}_{\sigma}$),
and long-range ($\epsilon^{\text{LR}}_{\sigma}$) contributions to $\epsilon_{\sigma}$, obtained by
considering only pairs separated by 0-15 \AA, 15-30 \AA, and more than 30 \AA, respectively.

\section*{Acknowledgments}
The authors thank H. Flechsig and M. D\"{u}ttmann for valuable discussions.
Y.D. is Postdoctoral Researcher at the Belgian Fund for Scientific Research (F.R.S.-FNRS),
and acknowledges support from the Walloon region through a WBI grant.

\section*{Author Contributions}
Y.D. designed and performed study. Y.D. and A.S.M analyzed data and wrote the paper.

\end{document}